\documentclass[12pt, a4paper]{article}
\usepackage{amsmath,amssymb,amsfonts}
\usepackage{graphicx}
\usepackage[colorlinks]{hyperref}

\begin{document}

\title{Induced vacuum current and magnetic field in the background of a vortex}

\author{Volodymyr M. Gorkavenko${}^{1,}$\thanks{E-mail: gorka@univ.kiev.ua},  Iryna V. Ivanchenko${}^{2,}$\thanks{E-mail: iiv.ivanchenko@gmail.com},\\
 Yurii A. Sitenko${}^{2,3,}$\thanks{E-mail: sitenko@itp.unibe.ch}\\
 \it \small ${}^{1}$Department of Physics, Taras Shevchenko National
 University of Kyiv,\\ \it \small 64 Volodymyrs'ka str., Kyiv
 01601, Ukraine\\\phantom{11111111111}\\
 \it \small ${}^{2}$Bogolyubov Institute for Theoretical Physics,
 \it \small National Academy of Sciences of Ukraine,\\
 \it \small 14-b Metrologichna str., Kyiv 03680,
 Ukraine\\\phantom{bhh}\\
  \it \small ${}^{3}$Institute for Theoretical Physics, University of
  Bern,\\
 \it \small Sidlerstrasse 5, CH-3012 Bern, Switzerland}
 \date{}

\maketitle

\begin{abstract}

A topological defect in the form of the Abrikosov-Nielsen-Olesen
vortex is considered as a gauge-flux-carrying tube that is
impenetrable for quantum matter. Charged scalar matter field is
quantized in the vortex background with the perfectly reflecting
(Dirichlet) boundary condition imposed at the side surface  of the
vortex. We show that a current circulating around the vortex and a
magnetic field directed along the vortex are induced in the vacuum,
if the Compton wavelength of the matter field exceeds considerably
the transverse size of the vortex. The vacuum current and magnetic
field are periodic in the value of the gauge flux of the vortex,
providing a quantum-field-theoretical manifestation of the
Aharonov-Bohm effect. The total flux of the induced vacuum magnetic
field attains noticeable finite values even for the Compton
wavelength of the matter field exceeding the transverse  size of the
vortex by just three  orders of magnitude.

%Keywords: {Vacuum polarization; vortex; current; magnetic flux.}
\end{abstract}

%\ccode{PACS numbers: 11.27.+d, 11.10.Kk, 11.15.Tk}

\section{Introduction}

Spontaneous breakdown of continuous symmetries gives rise to
topological defects (texture solitons) of various kinds. In
particular, if the first homotopy group of the group space of the
broken symmetry group is nontrivial, then a linear topological
defect known as the Abrikosov-Nielsen-Olesen (ANO) vortex
\cite{Abr,NO} is formed. The vortex is described classically in
terms of a spin-0 (Higgs) field which condenses and a spin-1 field
which corresponds to the spontaneously broken gauge group; the
former is coupled to the latter in the minimal way with constant
${\tilde e}_H$. The transverse size of the vortex is of the order of
the correlation length, $\hbar(m_H c)^{-1}$, where $m_H$ is the mass
of the condensate field. Single-valuedness of the condensate field
and finiteness of the vortex energy implement that the vortex flux
is related to ${\tilde e}_H$: $\Phi=\oint d \textbf{x}  \textbf{A}
(\textbf{x})=2\pi \hbar c {\tilde e}_H^{-1}$, where $ \textbf{A}
(\textbf{x})$ is the vector potential of the spin-1 gauge field, and
the integral is over a path enclosing the vortex tube once. The
quantized matter field is coupled minimally to the spin-1 field with
constant $\tilde e$; thus quantum effects in the background of the
ANO vortex depend on the value of $\tilde e \Phi$. The case of
${\tilde e}_H=2\tilde e$ $(\Phi=\pi \hbar c {\tilde e}^{-1}$, half
of the London flux quantum) is realized in ordinary
Bardeen-Cooper-Schrieffer superconductors where the Cooper-pair
field condenses and, in addition, there are normal electron
(pair-breaking) excitations, see \cite{Hue}; the cases of fractional
values of ${\tilde e}_H/(2\tilde e)$ can be realized in chiral
superfluids, liquid crystals and quantum liquids, see
\cite{Nel,Vol}.

An issue of ANO vortices under the name of cosmic strings is widely
discussed in the context of astrophysics and cosmology for more than
three decades \cite{vilenkin,hind}. The formation of such
topological defects during the cosmological expansion is predicted
in most interesting models of high energy physics, providing an
important link between cosmology and particle physics, see review in
\cite{Cop}. Cosmic strings serve as plausible sources of detectable
gravitational waves, high-energy cosmic rays and gamma-ray bursts
\cite{Ber,Bra,Jack}.

While considering the effect of the ANO vortices on the vacuum of
quantum matter, the following circumstance should be kept in mind:
the phase with broken symmetry exists outside the vortex and the
vacuum is to be defined only there; hence the quantum matter field
is not permitted to penetrate inside the vortex, obeying a boundary
condition at its side surface. Further, we shall assume that the
interaction between the ANO vortex and the quantum matter field is
mediated by the vector potential of the vortex-forming spin-1 field
only. The direct coupling between the vortex-forming spin-0 field
and the quantum  matter field can be neglected. The latter is
consistent with the requirement that the Compton wavelength of the
quantum matter field is much larger than the transverse size of the
vortex, and this requirement will be substantiated in the course of
the present study. Thus, the ANO vortex has no effect on the
surrounding matter in the framework of classical theory, and such an
effect, if exists, is of purely quantum nature. The effect should be
denoted as a quantum-field-theoretical manifestation of the famous
Aharonov-Bohm effect \cite{Aha}, see review \cite{Pesh}, and is
characterized by the periodic dependence on the value of the vortex
flux, $\Phi$, with the period equal to the London flux quantum,
$2\pi \hbar c {\tilde e}^{-1}$.

In the present paper, we shall study the current and the magnetic
field which are induced in the vacuum of the quantized charged
scalar matter field by the ANO vortex. These vacuum characteristics
were considered previously in the approximation neglecting the
transverse size of the vortex, see \cite{SitV,Jackiw} and references
therein. The aim of the present study is to take account for the
nonvanishing transverse size. We follow the lines of the works
\cite{newstring} -- \cite{newstring4} where the Casimir energy and
force in the background of the ANO vortex are studied. The quantized
matter field is assumed to vanish at the side surface of the vortex,
and the tension spread over the vortex is neglected; natural units
$\hbar=c=1$ will be used in the following.

\section{Induced vacuum current and total magnetic flux}

We start with Lagrangian for a complex scalar field $\psi$  in
$(d+1)$-dimensional space-time
\begin{equation}\label{0}
\mathcal{L}=({\mbox{ $\nabla$}}_\mu\psi)^*({\mbox{
$\nabla$}}^\mu\psi)-m^2\psi^*\psi,
\end{equation}
where ${\mbox{ $\nabla$}}_\mu$ is the covariant derivative and $m$
is the mass of the scalar field. The operator of a second-quantized
 scalar field can be represented in the form
\begin{equation}\label{1}
 \Psi(x^0,{\textbf{x}})=\sum\hspace{-1.4em}\int\limits_{\lambda}\frac1{\sqrt{2E_{\lambda}}}\left[e^{-{\rm i}E_{\lambda}x^0}\psi_{\lambda}({\bf x})\,a_{\lambda}+
  e^{{\rm i}E_{\lambda}x^0}\psi_{\lambda}^*({\bf
  x})\,b^\dag_{\lambda}\right];
\end{equation}
 $a^\dag_\lambda$ and $a_\lambda$ ($b^\dag_\lambda$ and
$b_\lambda$) are the scalar particle (antiparticle) creation and
destruction operators satisfying commutation relations
\begin{equation}\label{2}
[a_\lambda,a^+_{\lambda'}]_-=
[b_\lambda,b^+_{\lambda'}]_-=\langle\lambda'|\lambda\rangle;
\end{equation}
$\lambda$ is the set of parameters (quantum numbers) specifying the
state; wave functions $\psi_\lambda(\textbf{x})$ form a complete set
of solutions to the stationary Klein-Fock-Gordon equation
\begin{equation}\label{3}
 \left(-{\mbox{\boldmath $\nabla$}}^2  + m^2\right)  \psi_\lambda({\bf x})=E^2_\lambda\psi({\bf x}),
\end{equation}
 $E_\lambda=E_{-\lambda}>0$ is the energy of the state;
symbol $\sum\hspace{-1em}\int\limits_\lambda$ denotes summation over
discrete and integration (with a certain measure) over continuous
values of  $\lambda$.

In the present paper we are considering a static background in the
form of the cylindrically symmetric gauge flux tube of the finite
transverse size. The coordinate system is chosen in  such a way that
 the tube is along the $z$ axis.
  The tube in 3-dimensional space is obviously generalized to
 the $(d-2)$-tube in $d$-dimensional space by adding extra $d-3$
 dimensions as longitudinal ones.
 The covariant derivative is $\nabla_0=\partial_0$, $\mbox{\boldmath
$\nabla$}=\mbox{\boldmath $\partial$}-{\rm i} \tilde e\, {\bf V}$
with $\tilde e$ being the coupling constant of dimension
$m^{(3-d)/2}$ and the vector potential possessing only  one
nonvanishing component given by
\begin{equation}\label{4}
V_\varphi=\Phi/2\pi,
\end{equation}
outside the tube; here  $\Phi$ is the value of the gauge flux inside
the $(d-2)$-tube and $\varphi$ is the angle in  polar $(r,\varphi)$
coordinates on a plane which is transverse to the tube.   The
Dirichlet boundary condition at the side surface of the tube
$(r=r_0)$ is imposed on the scalar field:
\begin{equation}\label{5}
\left.\psi_\lambda\right|_{r=r_0}=0,
\end{equation}
i.e. the quantum matter is assumed to be perfectly reflected from
the thence impenetrable flux tube.

The solution to \eqref{3} and \eqref{5} outside the impenetrable
 tube of  radius $r_0$ takes form
\begin{equation}\label{6}
\psi_{kn{\bf p}}({\bf x})=(2\pi)^{(1-d)/2}e^{{\rm i}\bf{p
x}_{d-2}}e^{{\rm i}n\varphi}\Omega_{|n- {\tilde e}
\Phi/2\pi|}(kr,kr_0),
\end{equation}
where
\begin{equation}\label{7}
\Omega_\rho(u,v)=\frac{Y_{\rho}(v)J_{\rho}(u)-J_{\rho}(v)Y_{\rho}(u)}{\left[J^{2}_{\rho}(v)+Y^{2}_{\rho}(v)\right]^{1/2}},
\end{equation}
and $0<k<\infty$, $-\infty<p^j<\infty$ ($j=\overline{1,d-2}$), $n\in
\mathbb{Z}$ ($\mathbb{Z}$ is the set of integer numbers),
 $J_\rho(u)$ and $Y_\rho(u)$ are the Bessel functions of order $\rho$ of the first and  second
 kinds. Solutions \eqref{6} obey orthonormalization condition
\begin{equation}\label{8}
\int\limits_{r>r_0} d^{\,d}{\bf x}\, \psi_{kn{\bf p}}^*({\bf
x})\psi_{k'n'{\bf p}'}({\bf
x})=\frac{\delta(k-k')}{k}\,\delta_{n,n'}\,\delta^{d-2}(\bf{p}-\bf{p}').
\end{equation}

The vacuum current of scalar field is defined as
\begin{equation}\label{9}
\bold{ j}(\textbf{x})=\frac1{2{\rm i}}\left\langle {\rm vac}\left|
\left\{ [\Psi^+(x^0,\bold{x}),\mbox{\boldmath
$\nabla$}\Psi(x^0,\bold{x})]_+ -[\mbox{\boldmath
$\nabla$}\Psi^+(x^0,\bold{x}),\Psi(x^0,\bold{x})]_+ \right\}
  \right|{\rm vac}\right\rangle,
\end{equation}
with $[A, B]_+=AB+BA$. Using \eqref{1} and \eqref{6} we get
$j_r={\bf j}_{d-2}=0$ and
\begin{equation}\label{10}
j_\varphi(r)\equiv x^1 j^2(\textbf{ x})-x^2
j^1(\textbf{x})=(2\pi)^{1-d}\int d^{d-2}p\int\limits_0^\infty dk\,k
({\bf p}^2+k^2+m^2)^{-1/2}S(kr,kr_0),
\end{equation}
where
\begin{equation}\label{10a}
S(u,v)=\sum_{n\in\mathbb{Z}}\left(n-\frac{ \tilde
e\Phi}{2\pi}\right)\Omega_{|n- \tilde e\Phi/2\pi|}^2(u,v).
\end{equation}
Due to the infinite range of the summation, the last expression is
periodic in flux $\Phi$ with a period equal to $2\pi{\tilde
e}^{-1}$, i.e. it depends on quantity
\begin{equation}\label{11}
    F=\frac{ \tilde e\Phi}{2\pi}-\left[\!\left[\frac{\tilde e\Phi}{2\pi}\right]\!\right],
\end{equation}
where $[[u]]$ is the integer part of  quantity $u$ (i.e. the integer
which is less than or equal to $u$).

Let us rewrite \eqref{10a} in the form
\begin{equation}\label{11a}
S(u,v)=S_0(u)+S_1(u,v),
\end{equation}
 where
\begin{equation}\label{12}
S_0(u)=\sum_{n=0}^\infty\left[\left(n+1\!-\!F\right)J_{n+1-F}^2(u)\!-\!
\left(n\!+\!F\right)J_{n+F}^2(u)\right]
\end{equation}
and
\begin{equation}\label{15}
S_1(u,v)=\sum_{n=0}^\infty\left[(n+1-F)\Lambda_{n+1-F}(u,v)-(n+F)\Lambda_{n+F}(u,v)\right],
\end{equation}
where
\begin{equation}\label{16}
\Lambda_\rho(u,v)=\frac{J^{2}_{\rho}(v)\left[Y_\rho^2(u)-J_\rho^2(u)\right]-2J_\rho(v)J_\rho(u)Y_\rho(v)Y_\rho(u)}{Y^{2}_{\rho}(v)+J^{2}_{\rho}(v)}\,.
\end{equation}

Vacuum current $j_\varphi$ circulating around the $(d-2)$-tube leads
to the appearance of the vacuum magnetic field with strength
$B^{3...d}$ directed along the $(d-2)$-tube; this is a consequence
of the Maxwell equation
\begin{equation}\label{16a}
r\partial_r B^{3...d}_{(I)}({r})=- e j_\varphi({r}),
\end{equation}
where coupling constant $e$ differs in general from $\tilde e$. The
total flux of the induced vacuum magnetic field across a plane which
is orthogonal to the $(d-2)$-tube is defined as
\begin{equation}\label{16d}
  \Phi^{(I)}_d=2\pi\int_{r_0}^\infty dr\,r B^{3...d}_{(I)}({r})
\end{equation}
and is given by expression
\begin{equation}\label{16e}
  \Phi^{(I)}_d= e\pi\int\limits_{r_0}^\infty dr\,r j_\varphi(r)
  \left(1-\frac{r_0^2}{r^2}\right).
\end{equation}

Inserting $j_{\varphi}(r)$ \eqref{10} and changing the order of
integration over $r$ and ${\bf p}$, we obtain
\begin{equation}\label{a1}
  \Phi^{(I)}_d= e m^{d-3}\frac{(4\pi)^{(2-d)/2} }{2 \Gamma (d/2)}
  \int\limits_0^\infty \frac{du}{\sqrt{1+u^{2/(d-2)} }}\,
  \mathcal{D}(mr_0\, \sqrt{1+u^{2/(d-2)}}),
\end{equation}
where $\Gamma(v)$ is the Euler gamma-function and
\begin{equation}\label{a2}
  \mathcal{D}(y)=\int\limits_y^{\infty}dx \left(1-\frac{y^2}{x^2}\right)\int\limits_0^\infty
  \frac{dz\,z}{\sqrt{z^2+x^2}}\,S\left(z,z\frac{y}{x}\right).
\end{equation}
It should be noted that function $ \mathcal{D}(y)$ \eqref{a2} is
immediately related to the total induced vacuum magnetic flux in the
$d=2$ case:
\begin{equation}\label{a3}
\Phi^{(I)}_2=\frac{ e}{2m}\,\mathcal{D}(mr_0).
\end{equation}
Since $S_1(u,0)=0$, one can obtain
\begin{equation}\label{a4}
\mathcal{D}(0)=\int\limits_0^\infty dx
\int\limits_0^\infty\frac{dz\,z}{\sqrt{z^2+x^2}}\,
S_0(z)=\frac13\,F(1-F)\left(F-\frac12\right),
\end{equation}
and the total induced vacuum magnetic flux in the $d=2$ case is
finite in the limit of a singular (i.e. infinitely thin) vortex
filament, $r_0\rightarrow0$ \cite{SitB}:
\begin{equation}\label{a5}
\lim_{r_0\rightarrow0}\Phi^{(I)}_2=\frac{ e}{6m}\,
F(1-F)\left(F-\frac12\right).
\end{equation}

However, in the $d\geq 3$ cases the situation is different. One gets
\begin{equation}\label{a6}
\Phi^{(I)}_3=\frac{
e}{2\pi}\int\limits_0^\infty\frac{du}{\sqrt{1+u^2}}\,\mathcal{D}(mr_0\,\sqrt{1+u^2}),
\end{equation}
which yields \cite{SitV}
\begin{equation}\label{a7}
\Phi^{(I)}_3 = \hspace{-1.5em}\raisebox{-0.6em}{${}_{\small
r_0\rightarrow 0}$}-\frac{ e}{2\pi}\,\mathcal{D}(0)\ln (mr_0).
\end{equation}
In the $d>3$ cases one gets by changing the integration variable
\begin{equation}\label{a8}
\Phi^{(I)}_d= e
r_0^{3-d}\frac{(4\pi)^{(2-d)/2}}{\Gamma\left(\frac{d-2}{2}\right)}\int\limits_{mr_0}^\infty
dv\,(v^2-m^2r_0^2)^{(d-4)/2}\mathcal{D}(v),
\end{equation}
which yields
\begin{equation}\label{a9}
\Phi^{(I)}_d=\hspace{-1.5em}\raisebox{-0.6em}{${}_{\small
r_0\rightarrow 0}$} e r_0^{3-d}\frac{2
(2F-1)\sin(F\pi)\Gamma\left(\frac{d-1}{2}+F \right)
\Gamma\left(\frac{d+1}{2}-F
\right)}{d(d-1)(d-3)(4\pi)^{d/2}\Gamma(d/2)},\,\,d>3.
\end{equation}

\section{Numerical analysis  of the induced vacuum characteristics}

\begin{figure}[t]
\begin{center}
\includegraphics[width=85mm]{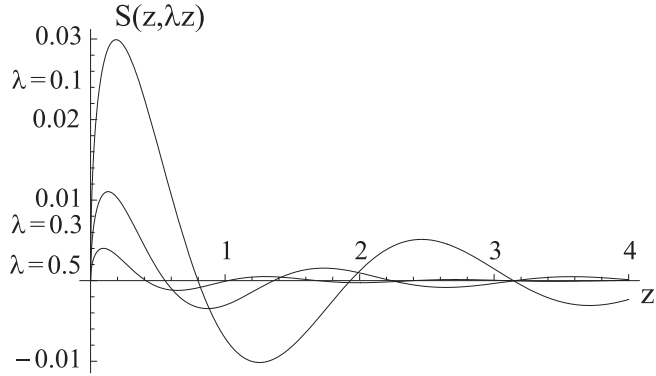}
\end{center}
\caption{Behavior of $S(z,\lambda z)$ at different values of
$\lambda$. \label{f1}}
\end{figure}

Let us rewrite expression \eqref{10} in the $d=2$ case in the
dimensionless form
\begin{equation}\label{17}
rj_\varphi(r)=\frac1{2\pi}\int\limits_0^\infty dz\,z
\left[z^2+\left(\frac{m
r_0}{\lambda}\right)^2\right]^{-1/2}S(z,\lambda z),
\end{equation}
where $\lambda=r_0/r$  $(\lambda\in [0,1])$. In the limit of a
singular  filament ($r_0=0$) expression \eqref{17} can be reduced to
the following form, see \cite{SitB},
\begin{multline}\label{s1}
rj_\varphi^{sing}(r)=\frac{\sin(F\pi)}{\pi^3}\times\\
\times\int\limits_{mr}^\infty
dw
\frac{w^2}{\sqrt{w^2-(mr)^2}}\left\{w\left[K^2_{1-F}(w)-K^2_F(w)\right]+(2F-1)K_F(w)K_{1-F}(w)
\right\}
\end{multline}
($K_\rho(u)$ is the Macdonald function of order $\rho$) with
asymptotics
\begin{equation}\label{s2}
rj_\varphi^{sing}(r)=e^{-1}rB^{sing}_{(I)}(r)=-\frac{\tan(F\pi)}{4\pi}\left(F-\frac12\right)^2+O\left[(mr)^2\right],\,\,
mr\ll 1
\end{equation}
and
\begin{equation}\label{s3}
rj_\varphi^{sing}(r)=2me^{-1}r^2B^{sing}_{(I)}(r)=\frac{\sin(F\pi)}{2\pi}\left(\!F-\frac12\right)\frac{e^{-2mr}}{\sqrt{\pi
mr}} \left(1+O\left[(mr)^{-1}\right]\right),\,\,mr\gg1,
\end{equation}
where $B^{sing}_{(I)}(r)$ is the vacuum magnetic field which is
induced in the $d=2$ case by a singular vortex filament. The total
induced vacuum magnetic flux in this case, see \eqref{a5}, attains
the maximal absolute value equal to $|e|/(72\sqrt{3}m)$ at $F=F_\pm
$, where
\begin{equation}\label{Flux}
  F_\pm=\frac12\left(1 \pm \frac1{\sqrt{3}}  \right).
\end{equation}

Our primary task is to compute numerically the induced vacuum
current in the $d=2$ case, see \eqref{17}, at $F=F_\pm$, when the
integral in \eqref{17} is likely to be most distinct from zero. It
can be shown that $S(z,\lambda z)$ is an oscillating function with
an amplitude that exponentially decreases at large $z$, see Fig.1.
So, for a vortex tube of nonvanishing radius, we have to compute
values of dimensionless quantity $r j_\varphi$ at different values
of $\lambda$. To do this, we perform high-precision numerical
integration in \eqref{17} with the help of a technique developed
earlier in \cite{newstring} -- \cite{newstring4} for the computation
of the vacuum energy density which is induced in the $d=2$ case by a
vortex tube of nonvanishing radius.The results can be approximated
by an interpolation function in the form
\begin{equation}\label{18}
rj_\varphi(r)=
\left[\frac{e^{-2x}}{\sqrt{x}}\right]\left[\left(\frac{x-x_0}{x}\right)^2\frac{P_3(x-x_0)}{x^3}\right]
\frac{Q_3(x^2)}{R_3(x^2)},\quad x>x_0,
\end{equation}
where $x=mr$, $x_0=mr_0$ and $P_n(y)$, $Q_n(y)$, $R_n(y)$  are
polynomials in $y$ of the $n$-th order with the $x_0$-dependent
coefficients. The first factor in the square brackets describes the
large distance behavior in the case of a zero-radius tube
(filament), the second factor in the square brackets is an
asymptotics at small distances from the side surface of the tube,
the last factor describes the behavior at intermediate distances.
Since the vortex tube is  impenetrable, $rj_\varphi(r)$ \eqref{18}
vanishes at $x\leq x_0$.

\begin{figure}[t]
\begin{center}
\includegraphics[width=155mm]{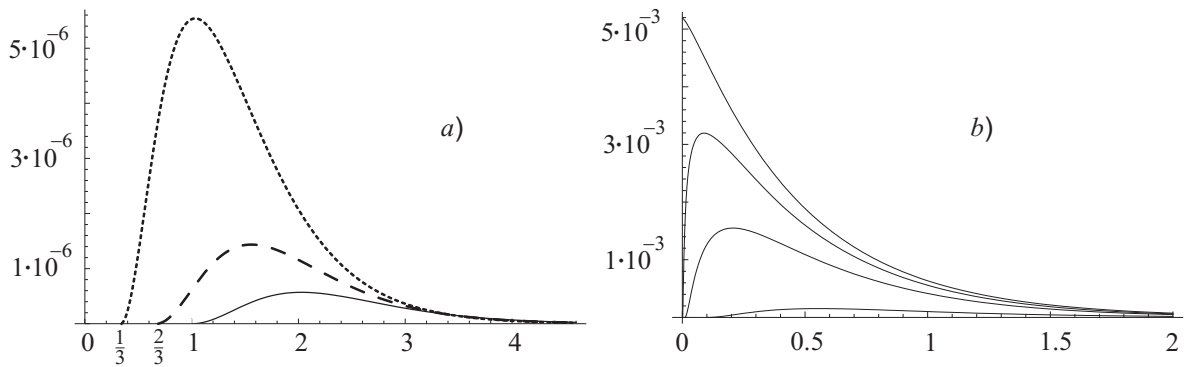}
\end{center}
\caption{The dimensionless induced vacuum current ($r j_{\varphi}$)
as a function of the dimensionless distance from the axis of the
tube $(x)$ for different values of the dimensionless tube radius
($x_0$): (a) solid line corresponds to $r j_{\varphi}\cdot 10^2$ for
$x_0=1$, dashed line corresponds to $r j_{\varphi}\cdot 10$ for
$x_0=2/3$ and dotted line corresponds to $r j_{\varphi}$ for
$x_0=1/3$; (b) the cases of a singular filament ($x_0=0$) and of a
tube with $x_0=10^{-3},10^{-2},10^{-1}$ are from up to down on the
plot. Variable $x$ is along the abscissa axis. \label{f2}}
\end{figure}

The results
 are presented on Fig.2. The
current is negligible for the  tube of large radius, i.e. of order
of the Compton wavelength and larger, $r_0\geq m^{-1}$, see Fig.2a.
But the current in the case of $r_0\ll m^{-1}$ is comparable with
the current in the case of a singular filament; note that the former
is always less in value than the latter, see Fig.2b.

Of particular interest is the behavior of the induced vacuum current
as the tube radius decreases. However, a direct numerical
computation in the case of $x_0<10^{-3}$ is a rather complicated
task, needing a long computational time. To surmount these
difficulties, one has to take account for the following two
circumstances. On the one hand, the case of a singular filament is
recovered as the tube radius tends to zero, $r_0\rightarrow0$. On
the other hand, in contrast to \eqref{s2}, the current in the case
of the nonvanishing tube radius vanishes quadratically in the
vicinity of the tube, see \eqref{18}; this is in accordance with the
analysis of the $r_0\rightarrow0$ limit for the solution to the
Klein-Fock-Gordon equation \cite{Bagrov, Gavrilov}. Therefore it is
reasonable to assume the following behavior at $r_0\ll m^{-1}$ and
$r-r_0\ll m^{-1}$:
\begin{equation}\label{s3}
r j_\varphi(r)=\left(\frac{r-r_0}{r}\right)^2 r j_\varphi^{sing}(r).
\end{equation}
The expected asymptotic behavior of the current in the case of the
extremely small tube radius ($r_0=10^{-9}m^{-1}$) is presented on
Fig.3.

\begin{figure}[t]
\begin{center}
\includegraphics[width=85mm]{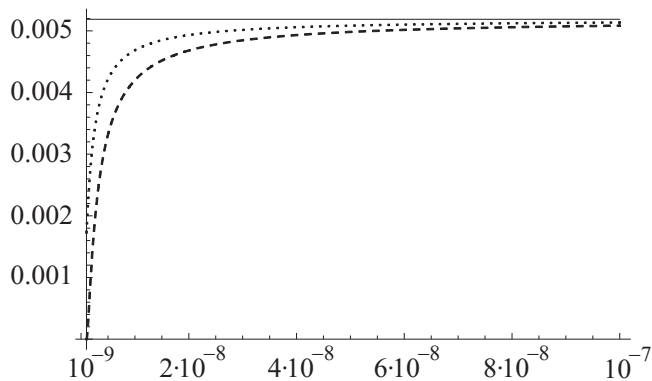}
\end{center}
\caption{The expected behavior of the dimensionless  induced current
($rj_\varphi$, dashed line) and the dimensionless induced magnetic
field ($e^{-1}r B_{(I)}$, dotted line) at small distances from the
side surface of the tube for the case of $x_0=10^{-9}$; solid line
corresponds to the case of a singular  filament ($x_0=0$), see
\eqref{s2}. The variable $x$ $(x>x_0)$ is along the abscissa
axis.\label{f3}}
\end{figure}

\begin{figure}[h]
\begin{center}
\includegraphics[width=145mm]{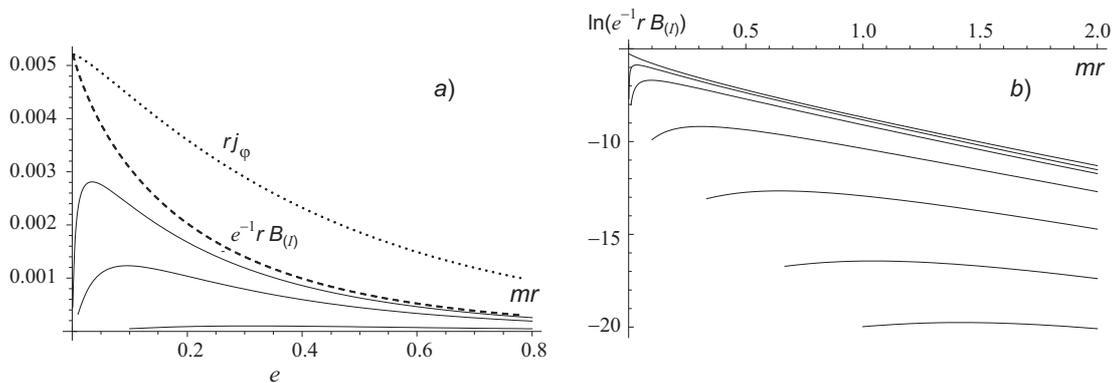}
\end{center}
\caption{The dimensionless  induced vacuum magnetic field $(e^{-1}r
B_{(I)})$  outside of tubes  of different radius: (a) the cases of a
singular filament (dotted and dashed lines) and tubes with
$x_0=10^{-3},10^{-2},10^{-1}$ are from up to down; (b) the case of a
singular  filament and  tubes with
$x_0=10^{-3},10^{-2},10^{-1},2/3,1/3,1$ in logarithmic scale are
from up to down. \label{f4}}
\end{figure}

Using \eqref{16a} and \eqref{18}, we compute numerically
$B_{(I)}(r)$, i.e. the induced vacuum magnetic field in the $d=2$
case. The results are presented on Fig.4 for the tube radius in the
range $10^{-3}m^{-1}\leq r_0\leq m^{-1}$. As the tube radius
decreases, the results approach to the result in the case of a
singular filament, excepting the region in the vicinity of the tube.
The expected asymptotic ($r\rightarrow r_0$) behavior of the induced
vacuum magnetic field in the case of the extremely small tube radius
($r_0=10^{-9}m^{-1}$) is presented on Fig.3.

Using \eqref{a1}, \eqref{a2}, \eqref{17} and \eqref{18}, we compute
numerically the total induced vacuum magnetic flux in the $d=2,3,4$
cases, i.e. $\Phi_2^{(I)}$, $\Phi_3^{(I)}$ and $\Phi_4^{(I)}$. The
results are presented on Fig.5 and in Table 1. In the $d=2$ case,
the absolute value of the flux induced by a filament is always
larger then the absolute value of the flux induced by a tube of the
nonvanishing radius. As the spatial dimension increases, the induced
flux becomes a more strongly decreasing function of the large tube
radius ($r_0\gtrsim m^{-1}$) and a more strongly increasing function
of the small tube radius ($r_0\ll m^{-1}$). Of particular interest
is the realistic case of $d=3$. Whereas the induced flux in the
unphysical case of a singular filament is infinite, see \eqref{a7}
and \eqref{a9} in the limit $r_0\rightarrow 0$, the induced flux in
the physical case of a tube of the nonvanishing radius is finite;
for instance, it attains a noticeable value of $0.0065\,e$ at
$r_0=0.001\,m^{-1}$ at $d=3$.

\begin{center}
\begin{tabular}{|c|c|c|c|c|c|c|}
\hline $x_0$ & 1 & 2/3 & $1/3$ & $10^{-1}$ & $10^{-2}$ & $10^{-3}$\\
\hline ${m \Phi^{(I)}_2}/e$ & $2.355\cdot 10^{-8}$ &
$5.762\cdot10^{-7}$ & $2.071\cdot10^{-5}$
& $4.878\cdot 10^{-4}$ & $3.657\cdot 10^{-3}$ & $6.217\cdot 10^{-3}$\\
\hline ${\Phi^{(I)}_3}/{ e} $ & $3.07\cdot 10^{-9}$ &
$8.829\cdot10^{-8}$ & $4.129\cdot10^{-6}$
& $1.483\cdot 10^{-4}$ & $2.223\cdot 10^{-3}$ & $6.547\cdot 10^{-3}$\\
\hline ${\Phi^{(I)}_4}/{(m e)} $ & $2.06\cdot 10^{-10}$ &
$7.06\cdot10^{-9}$ & $4.43\cdot10^{-7}$
& $2.675\cdot 10^{-5}$ & $1.254\cdot 10^{-3}$ & $1.585\cdot 10^{-2}$\\
\hline
\end{tabular}
\end{center}
{\small Tab.1 The dimensionless induced vacuum magnetic flux in
cases of dimension $d=2,3,4$  for tubes of  different radius.}

 %To be more precise induced magnetic flux by the magnetic tube of the very small radius $x_0\ll1$ can be estimated as
 %\begin{equation}\label{fluxtube}
 %  \frac{m}{2\pi e^2}\left( \Phi^{sing}-\Phi\right)\simeq -\frac{(1 - 2 F)^2 \tan(F \pi)}{16\pi} x_0.
 %\end{equation}

% Induced by the magnetic tube  flux is function of only tube thickness $x_0$. It can be generalized to the physically interesting case of $3+1$ space-time

\section{Summary}

\begin{figure}[t]
\begin{center}
\noindent\includegraphics[width=155mm]{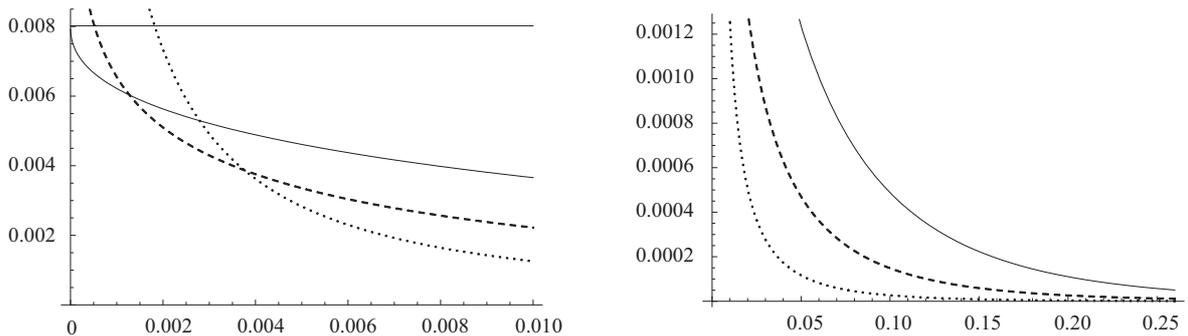}
\end{center}
\caption{The dimensionless  induced vacuum  magnetic flux in spaces
of different dimensionality as a function of the dimensionless tube
radius $(x_0)$:  $e^{-1}m \Phi^{(I)}_2$ -- solid line,
$e^{-1}\Phi^{(I)}_3$ -- dashed line,  $(em)^{-1}\Phi^{(I)}_4$ --
dotted line.  The case of small $x_0$ is  on the left,  and the case
of large $x_0$ is on the right. The case of $x_0=0$ and $d=2$ is
presented by a horizontal solid line on the left. \label{f5}}
\end{figure}

In the present paper, we consider the current and the magnetic field
which are induced in the vacuum of the quantized charged scalar
matter field by a topological defect in the form of the ANO vortex.
A perfectly reflecting (Dirichlet) boundary condition is imposed on
the matter field at the side surface of the vortex. The induced
current is circulating around the vortex, and the induced magnetic
field is directed along the vortex. Both the current and the
magnetic field are vanishingly small in the case of the vortex
transverse size being of the order of or exceeding the Compton
wavelength of the matter field (the dimensionless current is less
than $10^{-8}$, the dimensionless total flux of the magnetic field
is less then $10^{-9}$ in the $d=3$ case). Together with the results
of \cite{newstring,newstring4} about the Casimir force acting on the
vortex side surface, this confirms the conclusion that the vacuum
polarization of the quantized matter is almost absent in the case
when the mass of the Higgs field (forming the topological defect) is
of the order of or less than the mass of the matter field; the
vacuum polarization effects are essential for matter fields with
masses which are much smaller than a scale of the spontaneous
symmetry breaking (Higgs mass). As to the induced vacuum current and
magnetic field in the background of the ANO vortex, we show in the
present paper that they are essential in this latter case, being odd
in the value of the vortex flux, $\Phi$, and periodic in this value
with with the period equal the London flux quantum, $2\pi {\tilde
e}^{-1}$; to be more precise, they vanish at $F=0,1/2,1$, where $F$
is given by \eqref{11}, and are of opposite signs in the intervals
$0<F<1/2$ and $1/2<F<1$, with their absolute values being symmetric
with respect to the point $F=1/2$. The current and the magnetic
field decrease exponentially at large distances from the vortex,
while otherwise they behave similarly to the case of a singular
filament (vanishing transverse size) with the exception of a small
vicinity of the vortex tube. The latter distinction allows us to
eliminate an unphysical divergence which is present at $d\geq3$ for
the total induced vacuum magnetic flux in the case of a singular
filament. As long as the nonvanishing transverse size of the vortex
is taken into account, the total induced vacuum magnetic flux
becomes finite, attaining quite realistic values for the case of the
three-dimensional space even at the spontaneous symmetry breaking
scale exceeding the mass of the matter field by just three orders of
magnitude, see Fig.5 and Table 1. This can provide a possible
mechanism for generating primordial magnetic fields by cosmic
strings in early universe.

\section{Acknowledgments}

I.V.I. and Yu.A.S. acknowledge the support from the National Academy
of Science of Ukraine (project No.0112U000054).  The work of V.M.G.
was supported by the Swiss National Science Foundation grant SCOPE
IZ 7370-152581. The work of Yu.A.S. was supported by the
ICTP-SEENET-MTP grant PRJ-09 "Strings and Cosmology".

\end{document}